 \documentstyle[12pt]{article}
\newfont{\g}{eufm10}
\newcommand{\gtg}{\mbox{\g g}}
\newcommand{\gtsp}{\mbox{\g sp}}
\newcommand{\gtso}{\mbox{\g so}}
\newcommand{\gtpo}{\mbox{\g po}}

\newcommand{\gtb}{\mbox{\g b}}

\newcommand{\hgtg}{\mbox{$\hat{\gtg}$}}

\newcommand{\gta}{\mbox{\g a}}
\newcommand{\gtsl}{\mbox{\g sl}}
\newcommand{\gtgl}{\mbox{\g gl}}

\newcommand{\gtn}{\mbox{\g n}}

\newcommand{\nc} {\mbox{${\bf C}$}}

\newcommand{\nz} {\mbox{${\bf Z}$}}
\newcommand{\nr} {\mbox{${\bf R}$}}

\newcommand{\co}{\mbox{${\cal O}$}}

\newcommand{\cv}{\mbox{${\cal V}$}}

\newtheorem{theorem}{Theorem}[section]
\newtheorem{proposition}[theorem]{Proposition}

\newtheorem{example}[theorem]{Example}

\newtheorem{remark}[theorem]{Remark}
\newtheorem{remarks}[theorem]{Remarks}

\newtheorem{lemma}[theorem]{Lemma}

\title{{ \bf Universal Drinfeld-Sokolov Reduction and Matrices
of Complex Size}}
\author{Boris Khesin\thanks{Partially supported
by NSF grant DMS 9307086}
 and Feodor Malikov
\thanks{Partially supported
by NSF grant DMS 9401215}
\\
 Department of Mathematics, Yale University \\
 New Haven CT 06520 USA\\\\hep-th / 9405116}
\date{ }
\begin{document}
\baselineskip 20pt
\maketitle

\begin{abstract}
We construct affinization of the algebra $\gtgl_{\lambda}$ of
``complex size'' matrices, that contains the algebras $\hat{\gtgl_n}$
for integral values of the parameter.
The Drinfeld--Sokolov Hamiltonian reduction of
 the algebra $\hat{\gtgl_{\lambda}}$
results in the quadratic Gelfand--Dickey structure on the Poisson--Lie
group of all pseudodifferential operators of fractional order.

This construction is extended to the simultaneous deformation
of orthogonal and simplectic algebras which   produces self-adjoint
operators, and it has a counterpart for the
Toda lattices with fractional number of particles.
\end{abstract}

\section {{\bf Introduction}}

 As a rule  quadratic Poisson
 structures appear as either the Poisson
 bracket on a Poisson--Lie group or as a result of
 Hamiltonian reduction from the linear bracket on a dual Lie algebra.

This paper is devoted to a relation between
these two approaches to the classical \break
$W_n$--algebras (called also Adler--Gelfand--Dickey or higher
$KdV$-structures), natural infinite-dimensional
 quadratic Poisson structures on differential operators
 of $n^{th}$ order.

The noncommutative Hamiltonian
reduction (see\cite{Arn89Mat}, \cite{Mar-Wei}) for
the Gelfand--Dickey structures
(associated to any reductive Lie group)
is known as the reduction
 of Drinfeld and Sokolov (\cite{DriSok84Alg}).
They showed  that those quadratic
structures on $scalar \;\;
n^{th}\;\; order$ differential operators on the circle can be
obtained as a result of the two-step process
(restriction to a submanifold and taking quotient)
from the linear Poisson structure  on $\;matrix \;\; first\;\;
 order $ differential operators. The latter object is nothing
but the dual space to an affine Lie algebra on the circle (\cite{Fre},
\cite{Rei-Sem}).

On the other hand all Poisson $W_n$ algebras can be
regarded as Poisson submanifolds in a certain universal
Poisson--Lie group of pseudodifferential
 operators of arbitrary (complex) degree(\cite{KhesZakh93Poi}).
In such a way differential operators $DO_{n}=\{D^n +
u_1(x) D^{n-1} + u_2(x) D^{n-2}+...+u_n(x)\}$
for any $n$ turn out to be included as a Poisson submanifold
to a one-parameter family of pseudodifferential symbols
$\Psi DS_{\lambda}=\{D^{\lambda} +
u_1(x) D^{\lambda -1} + u_2(x) D^{\lambda -2}+...\}$.

At this point we bump into the following puzzle.
While a natural description of the Gelfand--Dickey
 structures through the above Poisson--Lie group
 exists for symbols of every complex degree $\lambda$ ,
 the Drinfeld--Sokolov
reduction is defined essentially for diffential
operators, that is for integer $\lambda=n$ and
first  $n$ coefficients $\{u_*(x)\}$.
The latter restriction is due to the very nature of
the Drinfeld--Sokolov reduction: it starts from
 the affine $\gtgl_n$ algebra and to find its counterpart
for complex $\lambda$ one needs to define groups of
 $\lambda\times\lambda$ matrices.

Actually the definition of $\gtgl_{\lambda},\lambda\in\nc$ has been known all
the time since representation theory of $\gtsl_{2}$ appeared.
It is simply the universal enveloping algebra of $\gtsl_{2}$ modulo the
relation: Casimir element is equal to $(\lambda-1)(\lambda+1)/2$. It was
Feigin, however, who placed this object in the proper context
(deformation theory) and applied it to calculation of the cohomology
of the algebra of differential operators on the line, see \cite{Fei88Alg}.

For technical reasons we have
here to replace the algebra $\gtgl_\lambda$
with its certain extension $\bar{\gtgl}_{\lambda}$. We
further construct an affinization of
the latter $\hat{\gtgl}_{\lambda}$.
This gives a family of algebras, $\lambda$ being the parameter,
such that for integral $\lambda$ the algebra has
a huge ideal and the corresponding quotient
is the conventional affine Lie alegebra $\hat{\gtgl_{n}}$. We prove the
 following conjecture of B.~Feigin and C.~Roger.

\begin{theorem}
The classical
Drinfeld--Sokolov  reduction  defined on $\hat{\gtgl_n}$
admits a one-parameter deformation to the Hamiltonian reduction
on $\hat{\gtgl_\lambda}$. As a Poisson manifold the result
of the reduction coincides
 with the entire
Poisson--Lie group of pseudodifferential operators equipped with
the quadratic Gelfand--Dickey structure.
\end{theorem}

It should be mentioned that, unlike integral $\lambda$ case,
for a generic $\lambda$ we can not use the formalism  of the
Miura transform (cf.\cite{KupWil81Mod}) or embedding of scalar higher order
 differential
operators into first order matrix ones by means of Frobenius
matrices. Both the operations are main tools in the classical
$\hat{\gtgl_n}$-case.

Feigin constructed also a simultaneous deformation of the
 symplectic and orthogonal algebras. We show that the corresponding
deformation of the Hamiltonian reduction results in the Gelfand--Dickey
bracket on self-adjoint pseudodifferential symbols.

We finish with a construction of the continuous
deformation of Toda lattice hierarchies.

The paper is essentially selfcontained. The
 Section 2 is devoted to basics in Poisson geometry and
Drinfeld--Sokolov reduction. Then we outline the
construction of the Poisson--Lie group of pseudodifferential
 operators and define the Adler--Gelfand--Dickey structures
 explicitly (Sect.3.1). Further we recall definition of
$\gtgl_\lambda$ and define its extension
and affinization  (Sect.3.2) which we beleive is
of interest by itself.
Remark that a similar interpolating object appears as a $sine$-algebra and
the algebra of ``quantum torus" (see \cite{G-L-O}).
In the Section 3.3
we construct the universal reduction of that algebra, resulting
in the structure on the Poisson --Lie group.
Section 4 is devoted to proofs.
 We conclude with discussion of $\gtsp$-, $\gtso$-cases of the reduction and
with consideration of a deformation of  the Toda lattices.

ACKNOWLEDGMENTS. We benefitted a lot from discussions with
many people. We are sincerely grateful to  Joseph Bernstein,
who explained to us a simple construction of the trace on
$\gtgl_{\lambda}$ (Sect. 3.2), to Claude Roger and Boris Feigin for
sharing with us the   conjecture, to Pavel Etingof, Laszlo Feher,
 Igor Frenkel,  Edward Frenkel, Vladimir Rubtsov,
and Gregg Zuckerman for numerous fruitful conversations.

\section {{\bf Poisson manifolds and Drinfeld-Sokolov reduction} }

\subsection{Preliminaries on Poisson geometry and Hamiltonian reduction}

\label{general-theory}

1. {\em Poisson algebras.} Symplectic
manifold is a pair $(M,\omega)$, where $M$ is a manifold
and
$\omega$ is a symplectic structure, i.e. non-degenerate skew-symmetric 2-form
on $M$. The symplectic form can be viewed as a non-degenerate
fiberwise linear map
\begin{equation}
\omega:\; TM\rightarrow T^{\ast}M,
\label{t-tstar}
\end{equation}
where $TM$ ($T^{\ast}M$) is a tangent (cotangent {\em resp.}) bundle
over $M$. For any $f,g\in C^{\infty}(M)$ set
\[\{f,g\}=dg(\Omega df).\]
where $\Omega=\omega^{-1} :T^{\ast}M\rightarrow TM$.
The bracket $\{.,.\}$  makes $C^{\infty}(M)$ a {\em Poisson algebra},
meaning
that the following holds:

\begin{eqnarray}
\mbox{the space}\;C^{\infty}(M) \mbox{ is a Lie
 algebra with respect to  }\{.,.\}
 \label {pois_alg_1}\\
\mbox{the Leibnitz identity}\;\{f,gh\}=\{f,g\}h+g\{f,h\}\;
 \mbox{fulfills}. \label{pois_alg_2}
\end{eqnarray}

 Let $Vect_{\omega}(M)$ be a Lie algebra of Hamiltonian vector fields
on $M$: \[Vect_{\omega}(M) =\{X\in Vect(M):\;X\omega=0\}.\]
Then the  map
$\Omega\circ d:\;C^{\infty}(M)\rightarrow Vect_{\omega}(M)$
is a Lie algebra homomorphism.

2. {\em Hamiltonian reduction.}  Let a Lie group $G$ act
 on a symplectic manifold $M$ by
symplectomorphisms, i.e. by diffeomorphisms preserving the symplectic
form $\omega$. Then there arises a morphisms of Lie algebras:
$\gtg\rightarrow Vect_{\omega}(M)$,
where $\gtg$ is a Lie algebra of $G$.

The action of $G$ on $M$ is called {\em
Poisson} if in addition the
latter morphism lifts to a Lie algebra morphism
$\gtg\rightarrow C^{\infty}(M).$
Denote by $H_{a}\in C^{\infty}(M),\;a\in \gtg,$ the image of $a$ in
$C^{\infty}(M)$, i.e. the Hamiltonian function corresponding to
an infinitesimal action $a$.

Denote by $\gtg^{\ast}$ a space dual to $\gtg$. The group $G$ naturally acts on
$\gtg^{\ast}$
(via coadjoint action).
For any Poisson action action of $G$ on $M$ there arises
a $G-$equivariant mapping called {\em momentum map}:
\[p:\;M\rightarrow\gtg^{\ast},\;<p(x),a>=H_{a}(x),\;\mbox{for}\; x\in
C^{\infty}(M),a\in\gtg.\]
Fix $a\in\gtg^{\ast}$ and denote by $G_{a}\subset G$ its
stabilizer. Obviously, the set $M_{a}=p^{-1}(a)$ is preserved by $G_{a}$.
Assume now that, first, $M_{a}$ is a manifold and that, second, so is
the quotient space
$F_{a}=M_{a}/G_{a}$. One can show that $F_{a}$ is a symplectic
manifold with respect to the symplectic form $\bar{\omega}$ defined by setting
\[\bar{\omega}(\bar{\xi} , \bar{\eta})=\omega(\xi,\eta),\]
where $\xi$ and $\eta$ are arbitrary preimages of
$\bar{\xi}$ and $\bar{\eta}$ with respect to  the natural projection
$TM_{a}\rightarrow TF_{a}$ (see \cite{Mar-Wei}).

The described passage from a symplectic manifold $M$ to a symplectic
manifold
$F_{a}$ of dimension less than dimension of $M$ by $2\,dimG_{a}$ is
called (noncommutative) {\em Hamiltonian reduction}

\begin{example} $T^{\ast}M$ is canonically a symplectic manifold
for any $M$.
 $T^{\ast}G$ is a symplectic manifold with a Poisson action of $G$ by
left translations. The momentum $T^{\ast}G\rightarrow
\gtg^{\ast}\approx T^{\ast}_{e}G$ is given by right translations to
the unit ($e$) of the group $G$.
The result of
the
hamiltonian reduction
with respect to the element $a\in\gtg$ is the orbit $\co_{a}$ of $a$
in the coadjoint representation equipped with the celebrated
Lie--Poisson--Kirillov--Kostant symplectic form.
\label{ex_co_orb}
\end{example}

3. {\em  Symplectic leaves.} We saw  above
 that for any symplectic manifold $M$ its algebra of
functions is a Poisson algebra. More generally,  $M$ is called a Poisson
manifold
if $C^{\infty}(M)$ is a Poisson algebra with respect to a certain
bracket $\{.,.\}$.

The bracket $\{.,.\}$ determines a Lie algebra morphism
$C^{\infty}(M)\rightarrow Vect(M)$.
It follows that $\{.,.\}$ can be regarded as a fiberwise linear map
$\Omega:\;T^{\ast}M\rightarrow TM$, where $\Omega$ already does not
necessarily
 come from a symplectic form (\ref{t-tstar}).
Corank of
restriction
of $\Omega$ to a fiber measures how far $\{.,.\}$ is from being induced by
a symplectic structure.
A notion of a symplectic form has the following substitute
for a generic Poisson manifold.

The assignment $M\ni x\mapsto \Omega(T^{\ast}_{x}M)\subset T_{x}M$ defines
a distribution on $M$. The integral submanifolds of this distribution
are called {\em symplectic leaves } of $M$. One shows that each symplectic
leaf is indeed a symplectic manifold (and, therefore, is also a Poisson
manifold). A Poisson submanifold is a manifold being a union of symplectic
leaves. The embedding of a Poisson submanifold of $M$
into $M$ is a Poisson morphism, meaning that the induced morphism of
algebras of functions is a morphism of Poisson algebras.

\begin{example} The dual space $\gtg^{\ast}$ is a Poisson manifold,
the bracket being defined by:
\[\{f,g\}(x)=<[d_{x}f,d_{x}g], x>,\]
where $f,g\in C^{\infty}(\gtg^{\ast})$, $d_{x}f$ signifies the value
of the differential of a function at the point $x$; therefore
$d_{x}f,\;d_{x}g\in \gtg$, and so the right hand side of the equality
is understood as a Lie bracket of a pair of elements of $\gtg$.
Symplectic leaves of $\gtg^{\ast}$ are exactly orbits of the coadjoint action,
see  Example~\ref{ex_co_orb}.
\label{dual_as_poiss}
\end{example}

Suppose a Lie group $G$ acts on a Poisson manifold $M$ by
diffeomorphisms
preserving the Poisson structure. Such an action is called {\em
Poisson }
if,
first, it preserves all symplectic leaves and, second, the induced Lie
algebra morphism of $\gtg$ to $Vect(M)$ lifts to a Lie algebra
morphism
$\gtg\rightarrow C^{\infty}(M)$.
In this case one can define a momentum $p:\;M\rightarrow \gtg^{\ast}$ so
that its restrictions to the symplectic leaves are exactly
 momenta of the above discussion. Assuming further that for some
$a\in\gtg^{\ast}$ , $M_{a}$ and $F_{a}=M_{a}/G_{a}$ are manifolds one
shows that $F_{a}$ is naturally a Poisson manifold and that its
symplectic leaves are exactly symplectic manifolds obtained via
Hamiltonian reduction applied to symplectic leaves of $M$.

\begin{example} Let $\gtn$ be a subalgebra of  $\gtg$ and $N$ be the Lie
group related to $\gtn$. A coadjoint action of $N$ on $\gtg^{\ast}$ is
Poisson,
the momentum being the natural projection $\gtg^{\ast}\rightarrow
\gtn^{\ast}$. So, $\gtg^{\ast}/N$ is a Poisson manifold.
\label{poissactondual}
\end{example}

An explicit calculation of the Poisson bracket of a pair
 of functions on $F_{a}$ can
be carried out as follows. Let $\pi: M_{a}\rightarrow F_a=M_{a}/G_{a}$ be
the natural projection. Let $f,g\in\nc^{\infty}(F_{a})$. Then $\pi^{\ast}f,\;
\pi^{\ast}g$ are functions on $M_{a}$. Choose an arbitrary continuation of each
of the functions on the entire $M$ and denote it also $\pi^{\ast}f,\;
\pi^{\ast}g$. Set
\begin{equation}
\label{def_poiss_quot}
\{f,g\}(\pi (x))=
\{\pi^{\ast}f,\pi^{\ast}g\}(x),\mbox{ for any $x\in M_{a}$}.
\end{equation}
What was said above is enough to prove that, despite
the
obvious ambiguities in this definition, the result is uniquely determined .

\subsection{Drinfeld-Sokolov construction}

\label{drinf-sok-constr}

 Drinfeld-Sokolov reduction is the procedure outlined in 2.1.3,
 especially in Example~\ref{poissactondual},
in the case when $\gtg$ is an affine Lie algebra and $\gtn$ is its
``nilpotent" subalgebra. To make a precise statement  let us fix the
following notations:

$\nc[z,z^{-1}]$ is a ring of Laurent polynomials, $\nc[[z]]$ is
a ring of formal
power series, $\nc((z))=\nc[z,z^{-1}]+\nc[[z]]$;

$\gtg=\gtgl_{n},\; \gtn\in \gtg$ is the subalgebra of
strictly
upper triangular matrices;

$\gta(z)=\gta\otimes\nc((z))$ for any Lie algebra $\gta$;
 $\gta(z)$
is called a {\em loop
algebra}; its elements can be  thought of as
``formal'' functions of $z\in\nc^{\ast}$
 with values in $\gta$;

$\hgtg = \gtg\oplus\nc$ is the corresponding affine Lie algebra, the universal
central extension of $\gtg$ by the
cocycle being given by
$\phi (f(z),g(z))=\, Res_{z=0}Tr(f(z)dg(z))$;

$A,\,N,\,G,\, A(z),\, N(z),\, G(z),\, \hat{G}$
are Lie groups
related to
 $\gta,\gtn, \gtg,\gta(z),\break\gtn(z), \gtg(z),\, \hgtg$.

The dual space $\gtg(z)^{\ast}$ is naturally isomorphic to
$\gtg(z)$ by
means of the invariant inner product (``Killing form'')
\[(f(z),g(z))=Res_{z=0}Trf(z)g(z)z^{-1}.\]

The dual space $\hgtg^{\ast}=\gtg((z))\oplus\nc$ can be  identified
with the space of 1st order linear differential operators on the circle
 with matrix
($n\times n$) coefficients $DO_{n\times n}$.

The correspondence $\hgtg^{\ast}\rightarrow DO_{n\times n}$
is established
by
\begin{equation}
\label{dual-diff}
\;(f(z),k)\mapsto
-k z\frac{d}{dz}+f(z).
\end{equation}

\begin{proposition}(\cite{Fre}, \cite{Rei-Sem})
The identification above makes the coadjoint action of
the group  $G[z,z^{-1}]$ on $\hgtg^{\ast}$  into
the gauge action on differential operators:

\[T(z)\cdot (f(z),k)=(-kzT(z)'T(z)^{-1} + T(z)f(z)T(z)^{-1},k),\]
here and elsewhere `` $'$ ''means the application of the operator $d/dz$.
\end{proposition}

\begin{remark}
The operators above can be viewed as differential operators on the circle
$z=\frac{1}{\sqrt{-1}}\exp(\sqrt{-1}\tau)$:

\begin{equation}
k \frac{d}{d\tau}+
f(\frac{1}{\sqrt{-1}}\exp(\sqrt{-1}\tau)).
\end{equation}
Solutions to differential equations with matrix coefficients are
vector functions. Natural action of $G[z,z^{-1}]$ on solutions
induces the gauge action
of $ G$ on  differential operators.
\end{remark}

Further we fix a hyperplane in $\hgtg$ by fixing a cocentral term:
 $\hgtg^{\ast}_{1}=\{(f(z),1),\;f(z)\in\gtg\}$. Obviously,
$\hgtg^{\ast}_{1}
\subset\hgtg^{\ast}$ is a Poisson submanifold: the Lie-Poisson bracket on
the dual space $\hgtg^{\ast}_{1}$ admits restriction to the hyperplane.

The coadjoint action of the
subgroup $\gtn$
on $\hgtg^{\ast}_{1}$ is Poisson. Consider its momentum map
\[p: \hgtg^{\ast}_{1}\rightarrow \gtn^{\ast}.\]
If the space $\gtn^{\ast}$ is identified with lower triangular matrices
then the momentum map $p$
 is nothing but the projection of (functions with values in) matrices onto
their lower-triangular parts.

To start Hamiltonian reduction we need to fix a point
 in the image of the momentum map. Regard the (lower triangular) matrix
\begin{equation}
\label{preim-fin}
\Lambda=\left( \begin{array}{cccccc}
0&0&0&\ldots&0&0\\
1&0&0&\ldots&0&0\\
0&1&0&\ldots&0&0\\
0&0&1&\ldots&0&0\\
.&.&.&.&.&.\\
0&0&0&\ldots&1&0 \end{array}
\right)
\end{equation}
as an element of $\gtn^{\ast}$.
The preimage $p^{-1}(\Lambda)$ is a manifold. It is, in fact, an affine
subspace:
$p^{-1}(\Lambda)=z\,d/dz+\Lambda+\gtb((z)),$
where $\gtb\in \gtg$ is the
subalgebra
of upper triangular matrices.

To perform the second step of the reduction we notice
that the quotient space \break
$p^{-1}(\Lambda)/N[z,z^{-1}]$ is also a  manifold.
Indeed, one can
 show that each $N[z,z^{-1}]-$orbit contains one and only one element of the
form
\[z\frac{d}{dz}+\left( \begin{array}{cccccc}
b_{1}(z)&b_{2}(z)&b_{3}(z)&\ldots&b_{n-1}(z)&b_{n}(z)\\
1&0&0&\ldots&0&0\\
0&1&0&\ldots&0&0\\
0&0&1&\ldots&0&0\\
.&.&.&.&.&.\\
0&0&0&\ldots&1&0 \end{array}
\right)\]

The space of first order differential operators of this form (
the corresponding  matrices are sometimes called  Frobenius matrices)
is in 1-1 correspondence with
(ordinary scalar)
differential operators $DO_{n}$ of order $n$  on the circle:
\[z\frac{d}{dz}+\left( \begin{array}{cccccc}
b_{1}(z)&b_{2}(z)&b_{3}(z)&\ldots&b_{n-1}(z)&b_{n}(z)\\
1&0&0&\ldots&0&0\\
0&1&0&\ldots&0&0\\
0&0&1&\ldots&0&0\\
.&.&.&.&.&.\\
0&0&0&\ldots&1&0 \end{array}
\right)\leftrightarrow \frac{d^n}{d\tau^n}+\bar{b}_{1}(\tau)
(\frac{d^{n-1}}{d\tau^{n-1}})+\cdots
+\bar{b}_{n}(\tau),\]
where $\bar{b}_k(\tau)=b_k(\exp({\sqrt{-1}\tau})/\sqrt{-1}),\;\tau\in\nr$.

The Hamiltonian reduction of  Sect.2.1.3 of the Kirillov--Kostant
structure on   $\hgtg^{\ast}_{1}$   equips
 $DO_{n}$ with a structure of a Poisson manifold.

On the other hand the space $DO_{n}$ is known to carry a Poisson
structure - the celebrated ``second Gelfand-Dickey structure''.

\begin{theorem} \cite{DriSok84Alg}
\label{th-dr-sok}
 The Poisson structure on $DO_{n}$ obtained as a
result of the  Hamiltonian reduction  described above is the second
Gelfand-Dickey
structure.
\end{theorem}

It is appropriate at this point to give a definition of the second
Gelfand-Dickey
bracket in the way Adler, Gelfand and Dickey did it, i.e. by explicit
formulas.
We will, however, refrain from doing so and instead describe a
generalization
of this construction to the case when $n$ is an arbitrary complex number.

\section{{\bf Differential operators of complex order and
 matrices of complex size.}}

\subsection{Poisson-Lie group of pseudodifferential symbols}

In this section we describe the main underlying structure,
 the Gelfand--Dickey bracket on the group of pseudodifferential symbols of
complex powers following \cite{KhesZakh93Poi}
(see also \cite{EnrKhorRad} for related questions).

Points of the Poisson manifold under consideration are
classical pseudodifferential
symbols, i.e. formal Laurent series of the following type:
$$
G=\{ D^{\lambda}+\sum_{k=-\infty}^{-1} u_k (z)D^{\lambda+k }
|u_k\in \nc((z)), \lambda \in \nc\}.
$$

This expression is to be understood as a convenient written form for
a semi-infinite sequence of functions $\{ u_k\}$.

This (infinite dimensional) manifold can be equipped with a group structure,
where product of  two such
symbols is a generalization of the Leibnitz rule $D\circ f(z)=f(z)D+f'(z)$
(that explains the meaning of the symbol $D=d/dz$). For  an arbitrary
(complex) power of $D$ one has:

\begin{equation}
\label{comm_d_lambda}
D^{\lambda}\circ f(z) = f(z)  D^{\lambda}+
\sum_{\ell \geq 1} \left(
{{\lambda}\atop \ell}\right) u^{(\ell )} (x) D^{{\lambda}-\ell}
\end{equation}
 where $\left( {{\lambda}\atop\ell}\right) =
{{\lambda}({\lambda}-1)\cdots ({\lambda}-\ell +1) \over \ell !}$.
The number $\lambda$ is called the order of a symbol.
It is easy to see that  every coefficient of the product of two symbols is
a differential polynomial  in coefficients of the factors.

{\bf Definition.}
The (quadratic generalized) Gelfand-Dickey Poisson
structure  on \break$G= \{ L=( 1+\sum_{k=-\infty}^{-1} u_k (x)D^k )
 D^{\lambda} \}$ is defined
as follows:

\medskip

 a) The value of the Poisson bracket of two functions at the given
point is determined by the restrictiong
 of these functions to the subset $\Psi DS_\lambda$ of symbols of fixed order
${\lambda}=const$.

\medskip

 b) The
subset ${\lambda}=const$ is an affine space, so we can identify the
tangent space to this subset with the set of operators of the form
$\delta L=( \sum_{k=-\infty}^{-1} \delta u_k D^k ) \circ D^{\lambda}$.

We can also identify the
cotangent space with the space of operators of the form
$X=D^{-\lambda} \circ DO $,
where $DO$ is a purely differential operator
 (i.e. polynomial in $D$) using the following pairing:

$$F_X (\delta L) := < X,\delta L> = Tr (\delta L\circ X) .$$
Here the product $\delta L\circ X$ is a symbol of an integer order
$\sum p_k(z) D^k$, and its
trace $Tr$ is defined as the residue at $z=0$ of $p_{-1}(z)$.

\medskip

 c) Now it is sufficient to define the bracket on linear functionals,
and

\begin{equation}
\label{scobka}
 \left\{ F_X ,F_Y \right\} \Bigl|_L =F_Y (V_{F_X} (L)),
\end{equation}
where $V_{F_X}$ is the following Hamiltonian mapping
$F_X \mapsto V_{F_X} (L)$
(from the cotangent space $\{X\}$ to
the tangent space $\{\delta L\}$):

\begin{equation}
\label{pole}
V_{F_X} (L)=(LX)_+ \ L-L(XL)_+ \ .
\end{equation}

\bigskip

\begin{remark}
Usually this definition is given only in the case when
$\lambda$ is a fixed positive integer and $L$ is a differential
 operator ($L_+=L$, here and above ``$+$'' means taking differential part
of a symbol of integral order), cf.\cite{Adl78Tra}, \cite{Dick}.
 The set $DO_{n}$
of purely differential operators
of order $n$ is a Poisson submanifold in the Poisson ``hyperplane''
$\Psi DS_{\lambda=n}$ of all
pseudodifferential symbols of the same order. Indeed,
for any operator ${L=D^n+u_{-1}(z)D^{n-1}+...
+u_{-n}(z)}$ and an arbitrary symbol $X=DO\circ D^{-n}$
the corresponding Hamiltonian vector
$V_{F_X} (L)=(LX)_+ \ L-L(XL)_+ \ $ is a differential operator of the order
$n-1$ and hence all Hamiltonian fields leave  the submanifold $DO_{n}$
of such operators ${L}$ invariant.

Exactly those Poisson submanifolds arise as a result of Hamiltonian reduction
in the Drinfeld-Sokolov construction.
For an arbitrary (noninteger) $\lambda$ one has no counterparts
of ``purely differential operators'' (what would be the differential part of
$u(z)D^{1/2}$ ?) and of the corresponding Poisson submanifolds.

\end{remark}

\medskip

As we noted, regardless of
$\lambda$, there is a natural homeomorphism between $\Psi DS_{\lambda}$ and
semi-infinite sequence of coefficients $\{u_i(z)\}$:

\begin{equation}
\Psi DS_{\lambda}\approx \prod_{i\geq 1}
\nc((z))
\label{alldolarehomeo}
\end{equation}

 The Poisson structure on the group induces a family
 $\{.,.\}_{\lambda}$ of Poisson structures on $\prod_{i\geq 1}
\nc((z))$, with polynomial dependence on $\lambda$ (combine
formulas (\ref{comm_d_lambda} -~\ref{pole})).

 The following filtration of the space
 $\prod_{i\geq 1}\nc((z))$ will be used later.

Represent it
as $\prod_{i=1}^{k} \nc((z))\times \prod_{i\geq k+1}\nc((z))$ and let
$i_{k}$ be the projection on the first factor. This gives an embedding of the
spaces of functions
\[i_{k}^{\ast}:\;Fun(\prod_{i=1}^{k} \nc((z)))\hookrightarrow
 Fun(\prod_{i\geq 1}\nc((z))).\]
Set $W_{k}=i_{k}^{\ast}(Fun(\prod_{i=1}^{k} \nc((z))))$. The
sequence $\{W_{k}\}$ forms the
 filtration
\begin{equation}
\label{filtrondiffop}
W_{1}\subset W_{2}\subset\cdots\subset
Fun(\prod_{i=1}^{k} \nc((z))),\;\cup_{i\geq 1}W_{i}=
Fun(\prod_{i=1}^{k} \nc((z))).
\end{equation}

\begin{proposition}
This filtration satisfies the condition:

\begin{equation}
\label{g_d_br-filtr}
\{W_{i},W_{j}\}_\lambda\subset W_{i+j}.
\end{equation}
\end{proposition}

{\em Proof.} Reformulating the statement
one needs to extract from the definition above
that for $L_0 \in \Psi DS_\lambda$ the bracket
$$
\{a(z)D^{i-1-\lambda},
\ b(z)D^{j-1-\lambda}\}_{\lambda}(L_0+L_1)
$$
does not depend on $L_1$ if  $deg\ L_1 \leq\lambda -i-j,
 \, L_0\in \Psi DS_\lambda$.

It follows from the explicit formulas
(\ref{scobka}-\ref{pole}) :
$$
{\frac{d}{d\epsilon}}\Bigl|_{\epsilon=0}\{A,B\}(L_0+\epsilon L_1) \break
 =Tr((L_0 A)_+ \ L_1 B-L_0(AL_1)_+ B +(L_1 A)_+ \ L_0 B-L_1(AL_0)_+ B).
$$
Calculation of the degrees shows that the right hand side vanishes
under the condition $deg\ L_1 \leq\lambda -i-j$. $\Box$

\bigskip

It should be mentioned that an analogous fact is basic
for ``quantum'' counterparts
of the Gelfand--Dickey Poisson structures (called quantum $W$--algebras).

\begin{remarks} (i) The Lie group structure of $G$ is
 compatible with the Gelfand--Dickey
structure and makes the group
into a Poisson--Lie one (see \cite{KhesZakh93Poi}).
Its Lie algebra $\gtg=\{\sum_{k=1}^{\infty} u_k(z)D^{-k}+\lambda\cdot\ log\,D
| u_k\in \nc[z,z^{-1}], \ \lambda \in \nc \}$ is a Lie bialgebra.
The formal expression $log\,D$ can be regarded as the velocity vector to the
one-parameter subgroup $D^\lambda$:
$$
d/d\lambda \Bigl|_{\lambda=0} \ D^{\lambda}
=log \,D\circ D^{\lambda} \Bigl|_{\lambda=0} =log \,D.
$$
and the commutation relation for the $log\ D$ and any symbol
 can be extracted from (\ref{comm_d_lambda})
$$
[log \,D, f(z)D^k] =\sum_{k\geq 1}^{} {(-1)^{k+1} \over k} \
f^{(k)}(z) D^{n-k}.
$$

The dual space to $\gtg$ is also a Lie bialgebra. It is nothing but the
unique central extension $\hat{DO}$ of the Lie algebra of all
differential operators
$DO=\{\sum_{j=1}^{n} a_j (z)D^j \}$ on the circle (\cite{KP}), known
also under the name
of $W_{1+\infty}$. The 2-cocycle
describing this extension can be given in terms of the outer derivation
$[log \,D, *]$:

\begin{equation}
c(A,B)=\int res \left( [log \,D,A]\circ B\right)
\end{equation}
where $A$ and $B$ are arbitrary
differential operators (see \cite{KhesKra91Cen}). The
restrictions of this cocycle to the subalgebra of vector fields
gives exactly the Gelfand--Fuchs cocycle
$$
c(u(z)D,v(z)D)=\frac{1}{6}\int u''(z)v(z)'dz
$$ which defines the Virasoro algebra.

(ii) Introduction of fractional power $D^{\lambda}$ is a particular
(Heisenberg algebra) case
 of formalism of fractional powers of  Lie algebra
generators, see \cite{Mal_F_F}. This formalism has been used for purposes of
representation theory.
It would be interesting to find its Poisson interpretation
 for other Lie algebras.

\medskip

\end{remarks}

\subsection{ Definition of
 $\gtgl_{\lambda},\;\lambda\in\nc$, and its extensions}
\label{alg_and_th_aff}

\subsubsection{ What is a matrix of complex size?}

Recall that $\gtsl_{2}$ is a Lie algebra on generators $e,h,f$ and
relations $[e,f]=h,\;[h,e]=2e,\;[h,f]=-2f$. There are different ways
to embed $\gtsl_{2}$ in $\gtgl_{n}$ and, hence, there are different
structures
of an $\gtsl_{2}-$module on $\gtgl_{n}$. Generically, however, the
structure of an $\gtsl_{2}-$module on $\gtgl_{n}$ is independent of the
embedding (see ~\cite{Kos})
and is given by
\begin{equation}
\gtgl_{n}=V_{1}\oplus V_{3}\oplus\cdots V_{2n-1},
\label{decompofgl}
\end{equation}
where $V_{i}$ stands for the irreducible $i-$dimensional
$\gtsl_{2}-$module. The image of a generic embedding of $\gtsl_{2}$ in
$\gtgl_{n}$
is called a principal $\gtsl_{2}-$triple. We do not discuss what
exactly the genericity condition is and confine to mentioning that an
example of a generic embedding is provided by sending
\begin{equation}
\label{ex_pr-sl2}
f\mapsto \left( \begin{array}{cccccc}
0&0&0&\ldots&0&0\\
1&0&0&\ldots&0&0\\
0&1&0&\ldots&0&0\\
0&0&1&\ldots&0&0\\
.&.&.&.&.&.\\
0&0&0&\ldots&1&0 \end{array}
\right)
\end{equation}
and continuing this map on the entire $\gtsl_{2}$.

In view of the decomposition (~\ref{decompofgl}) it is natural to ask
whether
the space $\oplus_{i\geq 0}V_{2i+1}$ admits a Lie algebra structure
consistent
with the existing structure of an $\gtsl_{2}-$module on it.
A construction of a 1-parameter family of such structures is as follows.

The universal enveloping algebra $U(\gtsl_{2})$ is a Lie algebra with
respect
to the operation $[a,b]=ab-ba$.
The element $C = ef +fe+\frac{1}{2}h^{2}$ generates
the center of $ U(\gtsl_{2})$.
The quotient
\[U(\gtsl_{2})/(C-\frac{1}{2}(\lambda-1)(\lambda+1))U(\gtsl_{2}),\;
\lambda\in\nc\]
is naturally a Lie algebra containing $\gtsl_{2}$. The fact that
its $\gtsl_{2}-$module structure is given by (~\ref{decompofgl})
is a consequence of much more general results of ~\cite{Kos2}. We point out
that in our case:

\begin{eqnarray}
\mbox{the elements $h^{i}e^{j},\;h^{i}f^{j}$ form a basis of the algebra, and}
\label{sl2decompofgll_basis}\\
\label{sl2decompofgll}
\mbox{the component $V_{2i+1}$ is generated
as an $\gtsl_{2}$-module by $e^{i}$,}
\end{eqnarray}

B.Feigin classified Lie algebra structures on
 $\oplus_{i\geq 0}V_{2i+1}$; in particular he
 proved that under a certain natural
assumption there are no families of Lie algebra structures on
 $\oplus_{i\geq 0}V_{2i+1}$ different from the
above mentioned, see ~\cite{Fei88Alg}.
 One proves (see Remark ~\ref{whysplits} below) that
if $\lambda$ is not integral the quotient is the sum of $\nc$ and a simple
(infinite dimensional) algebra, and
if $\lambda=\pm n, \;n\in\{1,2,...\}$ then
\[U(\gtsl_{2})/(C-\frac{1}{2}(\lambda-1)(\lambda+1))U(\gtsl_{2})\]
 contains an ideal and the quotient is isomorphic with
 $\gtgl_{n}$.
 For this reason  the algebra
\[U(\gtsl_{2})/(C-\frac{1}{2}(\lambda-1)(\lambda+1))U(\gtsl_{2})\]
is denoted
by $\gtgl_{\lambda}$ for an arbitrary complex $\lambda$.
Note that
our notations are
inconsistent in the sense that $\gtgl_{\lambda}$ is obviously different
from the conventional $\gtgl_{n}$ if $\lambda=n$. It is unfortunate but
 seems unavoidable; we will denote the finite-dimensional algebra by
 $\gtgl_{n}$ in the sequel.

\subsubsection {$\gtgl_{\infty}$ and  extensions of $\gtgl_{\lambda}$.}
Here we define 2 extensions of $\gtgl_{\lambda}$,
both being related to $\gtgl_{\infty}$ and one of them incorporating
a formal variable.

 Fix once and for all an infinite dimensional space $\cv$ with a
basis $\{v_{i},\;i=0,1,2,3,...\}$. This space carries a 1-parameter
family of $\gtsl_{2}-$module structures determined by:
\begin{equation}
\label{def_of_verma_mod}
fv_{i}=v_{i+1},\;hv_{i}=(\lambda-1-2i)v_{i},\;
ev_{i}=i(\lambda -i)v_{i-1},\;\lambda\in\nc.
\end{equation}
This, of course, makes $\cv$ into a Verma module $M(\lambda-1)$.
 Hence there
arises  the map
\begin{equation}
\label{sl2-gl_w_p_0}
U(\gtsl_{2})\rightarrow \gtgl_{\infty},
\end{equation}
where $\gtgl_{\infty}$ is the algebra of all linear
transformations of $\cv$.
Direct calculations show that $Cv_{i}=\frac{1}{2}(\lambda -1)(\lambda+1) v_{i}$
 for any $i$.
Therefore (\ref{sl2-gl_w_p_0}) factors through to the map
\begin{equation}
\label{sl2-gl_no_par}
\gtgl_{\lambda}\rightarrow \gtgl_{\infty}.
\end{equation}

\begin{lemma}
The map (\ref{sl2-gl_no_par}) is an embedding.
\end{lemma}
{\em Proof}. The map (\ref{sl2-gl_no_par}) is a morphism of $\gtsl_{2}-$
modules. Therefore it is enough to prove that each irreducible component of
 $\gtgl_{\lambda}$
is not annihilated by (\ref{sl2-gl_no_par}). But this is obvious: $V_{2i+1}$
is generated by $e^{i}$ (see (\ref{sl2decompofgll})) and $e^{i}$ is
a non-trivial operator on $M(\lambda-1)$. $\Box$

\medskip

 From now on we will identify $\gtgl_{\lambda}$ with its image in
 $\gtgl_{\infty}$.
The passage from a specific $\lambda$ to a formal parameter $t$ makes
(\ref{sl2-gl_w_p_0}) into the map
\begin{equation}
\label{sl2-gl}
U(\gtsl_{2})\rightarrow \gtgl_{\infty}\otimes\nc[t].
\end{equation}

$M(\lambda-1)$ is irreducible unless $\lambda=1,2,3,...$ and if the
latter condition is satisfied then it contains the unique proper
submodule $I_{\lambda}$
spanned by $v_{\lambda},v_{\lambda+1},...$  the corresponding
quotient being $V_{\lambda}$. This along with the definitions implies that
the image of $U(\gtsl_{2})$ under (~\ref{sl2-gl})
consists of matrices $A=(a_{ij}),\; i,j\geq 0$
  satisfying the
following
conditions:

(i) for any matrix $A=(a_{ij}),\;i,j\geq 0$ there exists a number $N$
such that $a_{ij}=0$ if $i>j+N$;

(ii) for any fixed $n$, $a_{i,i+n}$ is a polynomial in $i$;

(iii) if $\lambda=1,2,3,...$ then $a_{ij}(\lambda)=0$ once
$i<\lambda$ and
$j\geq\lambda$. (We naturally identify matrix elements with
polynomials in $t$.) In other words, in this case $A$ has the following
block form:

\[ \left(\begin{array}{cc}
B & 0\\
\ast & \ast \end{array}\right),\; B\in\gtgl_{n}. \]

\begin{remark}
\label{whysplits}
  The property (iii) explains why, under the integrality
condition $\lambda=n$, $\gtgl_{\lambda}$ ``contains'' the usual
$\gtgl_{n}$: $\gtgl_{n}\approx \gtgl_{\lambda}/J$, where
$J=\{A\in\gtgl_{\lambda}:\;Im(A)\subset I_{\lambda}\}$.

\end{remark}

Denote by $\gtgl$ the subalgebra of
$\gtgl_{\infty}\otimes\nc[t]$ satisfying the property (i) above and the
following weekened version of (ii) and (iii):

($\ast$) for any matrix $A\in\gtgl$ the properties (ii)
and (iii) can only be violated in a finite number of rows.

The algebra $\gtgl$ is one of the algebras we wanted to define. Definition
of the other is based on the following general notion which will be of use
later.

Let $W$ be a vector space and $A$  a subset of $ W\otimes\nc[t]$. The image of
 $A$ in $W$ under the evaluation map
 induced by projection $\nc[t]\rightarrow/(t-\epsilon)\nc[t]\approx\nc,\;
\epsilon
\in\nc$,
 will be denoted by $A_{\epsilon}$ and called {\em specialization}.

We now define $\bar{\gtgl}_{\lambda}$ to be a specialzitaion of $\gtgl$
 when $t=\lambda$. The following is an alternative
description of $\bar{\gtgl} _{\lambda}$
(it will not   be used in the sequel): $\bar{\gtgl} _{\lambda}$ is obtained
from $\gtgl _{\lambda}$ by, first, allowing infinite series of the form
$\sum_{i\geq 0}a_{i}e^{i},\; a_{i}\in\nc[h],$
(see (\ref{sl2decompofgll_basis})) and, second,
extending the result by the ideal of operators with finite-dimensional image.

\subsubsection{ Affinization and Coadjoint Representation of
 $\bar{\gtgl}_{\lambda}$}

1. {\em Trace.} The following simple and crucial construction was communicated
to us by J.Bernstein.  Observe that for any
$A=(a_{ij})\in\bar{\gtgl}_{\lambda}$ the sum
\[P(A,N)=\sum_{i=0}^{N-1}a_{ii}\]
is a polynomial in $N$ ( this is a consequence of $(\ast)$). Set
\begin{equation}
\label{def-tr}
Tr\,A=P(A,\lambda).
\end{equation}

It follows that both $\bar{\gtgl}_{\lambda}$ and
 the loop algebra
$\bar{\gtgl}_{\lambda}(z)=\bar{\gtgl}_{\lambda}\otimes\nc((z))$
carry an invariant non-degenerate    inner product defined by
\[(A,B)=Tr\,AB,\]
\[(A(z),B(z))=Res_{z=0}Tr\,A(z)B(z)z^{-1}dz\;(resp.).\]
Observe that the restriction of the trace to $\gtgl_{\lambda}$
and $\gtgl_{\lambda}(z)$ is
degenerate if $\lambda$ is a positive integer.

\medskip

2. {\em Affinization and coadjoint representation}
The loop algebra $\bar{\gtgl}_{\lambda}(z)$ admits a central extension
determined by the cocycle
\begin{equation}
\label{def-coc}
<A(z),B(z)>=Res_{z=0}\,Tr\, A(z)'B(z)\,dz.
\end{equation}
This provides the central extension  $\hat{\gtgl}_{\lambda}=
\bar{\gtgl}_{\lambda}(z)\oplus\nc\cdot c$.

 Using trace we make identifications
  $(\bar{\gtgl}_{\lambda})^{\ast}\approx
\bar{\gtgl}_{\lambda}$,
$(\bar{\gtgl}_{\lambda}(z))^{\ast}\approx
\bar{\gtgl}_{\lambda}(z)$, $(\hat{\gtgl}_{\lambda})^{\ast}\approx
\bar{\gtgl}_{\lambda}(z)\oplus\nc$ and extract subspaces
$(\hat{\gtgl_{\lambda}})^{\ast}_{k}\subset (\hat{\gtgl}_{\lambda})^{\ast},\;
k\in\nc$,
 where
$(\hat{\gtgl}_{\lambda})^{\ast}$ is all functionals equal to $k$ on the central
element $c$. The third
identification implies that elements of $(\hat{\gtgl}_{\lambda})^{\ast}_{k}$
are pairs $(A(z),k),\;A(z)\in\bar{\gtgl}_{\lambda}(z)$.
It follows from the definitions that
\begin{equation}
\label{coadj-act-at-point}
ad^{\ast}_{x(z)}(A(z),k))=([X(z),A(z)]-kzX(z)',k)
\end{equation}

\medskip

3. {\em Nilpotent subalgebra and subgroup.}
 Let $\gtn_{\lambda}\subset\bar{\gtgl}_{\lambda}$ be a subalgebra of strictly
upper triangular matrices and
 $\gtn_{\lambda}(z)$ the corresponding loop algebra.
Set $N_{\lambda}(z)=id \oplus \gtn_{\lambda}(z)$, where $id $ stands for the
identity operator.

\begin{lemma}
\label{exp-nilp}
$N_{\lambda}(z)$ is a group and the map
\[\exp:\;\gtn_{\lambda}(z)\rightarrow N_{\lambda}(z)\]
is a homeomorphism.
\end{lemma}

{\em Proof} is an easy exercise.

Exponentiating (~\ref{coadj-act-at-point}) one
 obtains that the coadjoint action
of the group $N(z)$  is given by

\begin{equation}
\label{coadj-act-group}
Ad^{\ast}_{X(z)}((A(z),k))=
(-zkX(z)'X(z)^{-1}+X(z)A(z)X(z)^{-1},k).
\end{equation}

\subsection{Drinfeld-Sokolov reduction on
$\hat{\gtgl}_{\lambda}$}

 The general theory (see sect.~\ref{general-theory}.3 and
Lemma \ref{coadj-act-group}) give
 the following:

(i) $(\hat{\gtgl}_{\lambda})^{\ast}$ and $(\gtn_{\lambda}(z))^{\ast}$ are
 Poisson
manifolds;

(ii) action of $N_{\lambda}(z)$ on $(\hat{\gtgl})^{\ast}$ is Poisson;

(iii)the natural projection (momentum)
\[p_{\lambda}:\;(\hat{\gtgl}_{\lambda})^{\ast}\rightarrow
(\gtn_{\lambda}(z))^{\ast},\]
 is Poisson and $N_{\lambda}(z)$-equivariant.

Analogously to what we did above (see sect.~\ref{drinf-sok-constr}),
cosider the matrix

\[f=\left( \begin{array}{cccccc}
0&0&0&\ldots&0&0\\
1&0&0&\ldots&0&0\\
0&1&0&\ldots&0&0\\
0&0&1&\ldots&0&0\\
.&.&.&.&.&.\\
 \end{array}
\right)\]
as an element of $(\gtn_{\lambda}(z))^{\ast}$.
( To justify the notation observe that from the $\gtsl_{2}-$point of view the
matrix above is simply the image of $f\in\gtsl_{2}$
in $\gtgl_{\infty}$.)
Restrict the momentum $p_{\lambda}$ to $(\hat{\gtgl}_{\lambda})^{\ast}_{1}$.
It is obvious that,
firstly, $p_{\lambda}^{-1}(f) \approx f+\gtb_{\lambda}(z)$, where
$\gtb_{\lambda}(z)$ is
the sublagebra of uppertriangular matrices, and, secondly, that
$N_{\lambda}(z)$
is the stabilizer of $f$. Hence there arises the quotient space
$p_{\lambda}^{-1}(f)_{\lambda}/N_{\lambda}(z),\;\lambda
\in\nc.$

\begin{proposition}
\label{action_group}

(i) Each $N_{\lambda}(z)-$orbit in
$p_{\lambda}^{-1}(f)$ contains one and only one Frobenius matrix,
i.e. an element of the
form
\[\left( \begin{array}{cccccc}
b_{1}(z)&b_{2}(z)&b_{3}(z)&\ldots\\
1&0&0&\ldots\\
0&1&0&\ldots\\
0&0&1&\ldots\\
.&.&.&.\end{array}
\right).\]

(ii) Elements of $N_{\lambda}(z)$ have no fixed points on
$p_{\lambda}^{-1}(f)$.

(iii) For any $\lambda$ the quotient space
$p_{\lambda}^{-1}(f)/N_{\lambda}(z)$ is isomorphic to the direct product
$\prod_{i\geq 1}\nc((z))$ equipped with the topology of projective
limit.
\end{proposition}

Again the general theory says that $p_{\lambda}^{-1}(f)/N_{\lambda}(z)$
 is a
Poisson manifold with the Poisson structure being reduced from the
Kirillov--Kostant structure on $(\hat{\gtgl}_{\lambda})^{\ast}$.
 It is naturally isomorphic to $\Psi DS_{\lambda}$ as a topological
space, the isomorphism being independent of $\lambda$ (c.f.

\begin{theorem} The spaces $p^{-1}_{\lambda}(f)/N_{\lambda}(z)$ (equipped
with the reduced Poisson structure)
 and $\Psi DS_{\lambda}$
(equipped with the quadratic Gelfand--Dickey structure)
are
isomorphic as Poisson manifolds for any $\lambda$.
\label{main_th}
\end{theorem}

\begin{remark}
 The (finite dimensional) $\gtgl_n$-quotient of the algebra
$\gtgl_{\lambda}$
(for integral $\lambda=n$) over the maximal ideal $J$
corresponds
precisely (via affinization and the classical Drinfeld--Sokolov construction)
to Poisson submanifolds $DO_n$
of purely differential operators
in the affine space $\Psi DS_{\lambda}$.
Indeed, functions vanishing on a Poisson submanifold
 form an ideal in the Lie algebra of functions
on the entire Poisson manifold. The corresponding quotient is nothing else but
the Poisson algebra
of functions on the  submanifold.
\end{remark}

\begin{remark}
 The first (linear) Gelfand--Dickey structure is defined by the formula
$V_A(L)=(LA-AL)_+$. Unlike the second (quadratic) structure above
the first one exists not on the
entire group of  $\Psi DS$, but only on the subspaces of
integral degree $\lambda$ (\cite{DriSok84Alg}).
Drinfeld--Sokolov reduction represents the linear Poisson structure on
scalar differential operators of $n^{th}$ order
as the reduction of a constant Poisson structure on $\hat{\gtgl}^*_n$
(i.e. on first order matrix differential operators).
This constant Poisson structure on the dual space is obtained by the freezed
argument principle applied to the Kirillov--Kostant structure at the point
$(0,  e_{1  n}) \in \hat{gl}^*_n$. Here 0 is the
coefficient at $z\frac{d}{dz}$, and $e_{1  n}$ is the  current on $S^1$
with the only nonvanishing entry equal 1 at (1,n)-place.

One can literally repeat  the arguments for the Hamiltonian reduction
from $\hat{\gtgl}_{\lambda}$. Then the finite matrix $e_{1 \ n}$ is to be
replaced by an infinite matrix, an element of $\hat{\gtgl}_{\lambda}^*$
 with the only
nonvanishing entry at the same place (1,n). This entry is singled
 out by the block
structure of $\gtgl_{\lambda}$ for integer $\lambda=n$. However,
for a generic ${\lambda}$ no such element is specified, and no
linear Poisson structure exists on the spaces $\Psi DS_{\lambda}$
after reduction.
\end{remark}

\section{ Proofs}
\subsection{ Affinization and Coadjoint Representation of $\gtgl$}

Proposition~\ref{action_group} says what canonical form of a matrix under the
action of the group $N_{\lambda}(z)$ is.
In our case, as it sometimes happens, it is easier
to find a canonical form of a
family of matrices than to do so with a single matrix.
In order to realize this
program we need to extend some of the above introduced notions to the case of
the algebra $\gtgl$.

\subsubsection{Affinization}

The algebra $\gtgl$ (incorporating the
 formal variable $t$) admits the trace: observe that for any $A=(a_{ij})$
the expression \[P(A,N)=\sum_{i=0}^{N-1}a_{ii}\]
is a polynomial on $N$ and set
\[Tr\,A=P(A,t).\]
It follows that both $\gtgl$ and
 the loop algebra
$\gtgl(z)=\gtgl\otimes\nc((z))$
carry an invariant non-degenerate $\nc[t]-$valued   inner product defined by
\[(A,B)=Tr\,AB,\]
\[(A(z),B(z))=Res_{z=0}Tr\,A(z)B(z)z^{-1}dz\;(resp.).\]

The loop algebra $\gtgl(z)$ admits the central extension
determined by the cocycle
\begin{equation}
\label{def-coc-2}
<A(z),B(z)>=Res_{z=0}\,Tr\, A(z)'B(z)\,dz.
\end{equation}
This provides the central extension  $\hat{\gtgl}=
\gtgl(z)\oplus\nc[t]$.

\subsubsection{Coadjoint Representation}

 Let as
usual $\nc[[t]]$ be the ring of formal power series,
$\nc((t^{-1}))
=\nc[t,t^{-1}]+\nc[[t^{-1}]]$.
$\nc((t^{-1}))$  is a $\nc[t]$-module and $\nc[t]\subset \nc((t^{-1}))$ its
$\nc[t]$-submodule. We identify $t^{-1}\nc [[t^{-1}]]$
with the quotient
$\nc((t^{-1}))/\nc[t]$. This equips $t^{-1}\nc [[t^{-1}]]$ with a
$\nc[t]$-module structure coming from $\nc((t^{-1}))/\nc[t]$.

 Existence of a nondegenerate invariant $\nc[t]$-valued
 inner product on $\gtgl$ gives that
\begin{equation}
\label{descript_of_dual}
\gtgl^{\ast}\approx t^{-1}\nc[[t^{-1}]]/\nc[t]\otimes_{\nc[t]}\gtgl.
\end{equation}

 The isomorphism is established by
 assigning to the pair \newline
 $(g(t),A(t)),\;g(t)\in t^{-1}\nc [[t^{-1}]],
A(t)\in\gtgl$ a functional by the formula
\[<g(t)A(t),B(t)>=Res_{t=0}g(t)TrA(t)B(t).\]
Similarly,
$\gtgl(z)^{\ast}\approx t^{-1}\nc [[t^{-1}]]\otimes_{\nc[t]}\gtgl((z))$
and
$(\hat{\gtgl})^{\ast}\approx t^{-1}\nc [[t^{-1}]]
\otimes_{\nc[t]}\gtgl((z))\oplus t^{-1}\nc [[t^{-1}]]$,
where element $(0,g(t))$ sends $(A(t,z),h(t))$ to
$Res_{t=0}g(t)h(t)$.

 For any $g(t)\in t^{-1}\nc[[t^{-1}]]$ set
\[(\hat{\gtgl})^{\ast}_{g(t)}=\{(g(t)A(t,z),g(t)),\;A(t,z)\in\gtgl((z)).\]

It is tempting to say that the dual space
$\hat{\gtgl}^{\ast}_{g(t)}$ for a fixed  $g(t)$ is in one-to-one
correspondence with $\gtgl((z))$. At least there is a map
\begin{equation}
\label{fromgl-to-hatgl}
\gtgl((z))\rightarrow \hat{\gtgl})^{\ast}_{g(t)},\;
A(t,z)\mapsto (g(t)A(t,z),g(t)).
\end{equation}

Properties of this map, however, essentially depend on the properties of
$g(t)$.
 Call an element of $t^{-1}\nc [[t^{-1}]]$ {\em rational} if it is equal to
Laurent expansion at $\infty$ of a rational function of $t$; otherwise
 an element of $t^{-1}\nc [[t^{-1}]]$ is called {\em irrational} .

\begin{lemma}
\label{descr-of-dual-precise}

(i) If $g(t)$ is irrational then
the map (~\ref{fromgl-to-hatgl}) is a one-to-one correspondence.

(ii) Let $g(t)=p(t)/q(t)$ for
some mutually prime $p(t),q(t)\in\nc[t].$ Then the map is a surjection
with ``kernel'' equal to the set of all matrices with entries divisible
by $q(t)$.
\end{lemma}

{\em Proof.} View elements of $(\hat{\gtgl})^{\ast}_{g(t)}$ as matrices with
coefficients in
 $g(t)\nc[t]/(g(t)\nc[t]\cap\nc[t])$
( see (~\ref{descript_of_dual})). Such a matrix determines the zero
functional if and only if all its entries are equal to 0, Lemma follows. $\Box$

\medskip

 The definitions imply that the coadjoint action of $\gtgl(z)$
preserves affine subspaces $(\hat{\gtgl})^{\ast}_{g(t)}$.
 Lemma~\ref{descr-of-dual-precise} implies that
the space $(\hat{\gtgl})^{\ast}_{g(t)}$ is always identified with
 $\gtgl((z))$ in the sense that
 in the case (ii) elements of $\gtgl((z))$ have to be viewed as matrices
with entries in the quotient ring $\nc[t]/q(t)\nc[t]$. Having
this in mind one obtains that
\begin{equation}
\label{coadj_act_of_alg}
ad^{\ast}_{X(t,z)}(A(t,z))=[X(t,z),A(t,z)]-zX(t,z)',\;A(t,z)\in
(\hat{\gtgl})^{\ast}_{g(t)}.
\end{equation}

The specialization map $\hat{\gtgl}\rightarrow\hat{\gtgl}_{\lambda}$
induces  emebddings
\[(\hat{\gtgl}_{\lambda})^{\ast}_{k}\hookrightarrow
(\hat{\gtgl})^{\ast}.\]
Direct calculations show that in fact

 \begin{equation}
\label{emb_of_duals}
(\hat{\gtgl}_{\lambda})^{\ast}_{1}\hookrightarrow
(\hat{\gtgl})^{\ast}_{1/(t-\lambda)},
\end{equation}
where $1/(t-\lambda)$
is viewed as a series $\sum_{i\geq 0}\lambda^{i}/t^{i+1}$.

The left dual to (~\ref{emb_of_duals}) is as follows
\begin{equation}
\label{left_inverse}
(\frac{1}{t-\lambda}A(t,z),  \frac{1}{t-\lambda})\mapsto A(\lambda,z).
\end{equation}

Indeed, Lemma~\ref{descr-of-dual-precise} implies that
$\hat{\gtgl}^{\ast}_{1/t-\lambda}$ is in  one-to-one correspondence with
$\gtgl((z))$ modulo the relation $t\approx \lambda$; this produces the desired
evaluation map.

 In exactly the same way
as in the $\bar{\gtgl}_{\lambda}-$case
one defines the nilpotent subalgebra $\gtn\subset\gtgl$,
  the corresponding loop algebra $\gtn(z)$, the corresponding group
 $N(z)=id \oplus \gtn(z)$ and proves that this group is exponential.

Exponentiating (~\ref{coadj_act_of_alg}) one obtains that the coadjoint action
of the group $N(z)$ is given by

\begin{equation}
\label{coadj-act-group-1}
Ad^{\ast}_{X(t,z)}(A(t,z))=-zX'X^{-1}+XA(t,x)X^{-1},\;X\in N[z,z^{-1}],\,
 A(t,x)\in
(\hat{\gtgl})_{g(t)},
\end{equation}
where, as always, if the assumption of Lemma~\ref{descr-of-dual-precise} (ii)
is satisfied, then all matrix entries are considered modulo the relation
$q(t)\approx 0$. In particular, when $q(t)=t-\lambda$ one obtains the coadjoint
action of $N_{\lambda}[z,z^{-1}]$ on $(\hat{\gtgl}_{\lambda})^{\ast}_{1}$.
This also means that the embedding (~\ref{emb_of_duals}) is
$N(z)$-equivariant, where $N(z)$ operates on
$(\hat{\gtgl}_{\lambda})^{\ast}_{1}$ via the evaluation map
$N(z)\rightarrow N_{\lambda}(z)$

\subsubsection { Conversion of a matrix to  the Frobenius form -- Proof
of  Proposition 3.7}
Fix $g(t)\in t^{-1}\nc[[t^{-1}]]$.
Consider the natural projection
\[p: (\hat{\gtgl})^{\ast}\rightarrow (\gtn(z))^{\ast}\]
and
denote its restriction  to the subspace $\hat{\gtgl}^{\ast}_{g(t)}$
by $p_{g(t)}$.
As above,
cosider the matrix

\[f=\left( \begin{array}{cccccc}
0&0&0&\ldots&0&0\\
1&0&0&\ldots&0&0\\
0&1&0&\ldots&0&0\\
0&0&1&\ldots&0&0\\
.&.&.&.&.&.\\
 \end{array}
\right)\]
as an element of $(\gtn(z))^{\ast}_{g(t)}$. The following is a natural
 generalization
of Proposition~\ref{action_group}.

\begin{proposition}
\label{action_group_1}

(i) Each $N(z)-$orbit in
$p_{g(t)}^{-1}(f)$ contains one and only one Frobenius matrix
\[\left( \begin{array}{cccccc}
b_{1}(t,z)&b_{2}(t,z)&b_{3}(t,z)&\ldots\\
1&0&0&\ldots\\
0&1&0&\ldots\\
0&0&1&\ldots\\
.&.&.&.\end{array}
\right),\]
where if $g(t)$ is rational then $b_{i}(t,z)$ is understood as an element
of an appropriate quotient ring.

(ii) Elements of $N(z)$ have no fixed points on $p^{-1}_{g(t)}(f)$.
\end{proposition}

Items (i) and (ii) of Proposition~\ref{action_group} follow from
Proposition~\ref{action_group_1} as an easy consequence of properties
of maps (~\ref{emb_of_duals},~\ref{left_inverse}). Item (iii) follow
from (i) and (ii) because all Frobenius matrices by definition belong to
$p^{-1}_{\lambda}(f)$.
The rest of this section is devoted to proving
Proposition~\ref{action_group_1}.

Suppose for simplicity that $g(t)$ is irrational.
Element $X\in N(z)$ converts $A\in \gtgl(z)$ to the Frobenius form if
 and only if
 the following equation
holds
\begin{equation}
\label{main_eq}
-zX'+XA=BX
\end{equation}
for some  Frobenius matrix $B$,
see (~\ref{coadj-act-group-1}). We will show that for any $A
\in\gtgl(z)$, the equation (~\ref{main_eq}) can be
effectively solved for unknown $X$ and $B$ and that
the solution is unique. This is achieved by the following
recurrent ``diagonalwise'' process.

Let $X=(x_{ij}(t,z)),\;A=(a_{ij}(t,z))$ and $B$ be as above.
(Note that all matrix entries
are ``functions'' of $z$ and $t$.) By definition, $x_{ii}(t,z)=1$. Therefore
(~\ref{main_eq}) for diagonal entries gives

\[x_{01}(t,z)=-b_{1}(t,z)+a_{00}(t,z),\;
 a_{i\,i}(t,z)+x_{i\,i+1}(t,z)=x_{i-1\,i}(t,z),\;i\geq 1.\]

This implies that
\[x_{i\,i+1}(t,z)=-\sum_{j=0}^{i}a_{j\,j}(t,z)-b_{1}(t,z),\;i\geq 1.\]
The condition $(\ast)$ in the definition of $\gtgl$ means that
$x_{n-1\,n}(n,z)=0$ for all sufficiently large positive integers $n$. So,
\[b_{1}(n,z)= \sum_{j=0}^{n-1}a_{jj}(t,z).\]
But the sum in the last
 expression is a polynomial in $n$, due to the definition
of $\gtgl$ , and
 this uniquely determines $b_{1}(t,z)$ as a polynomial in $t$.

Suppose we have found $b_{1}(t,z),\ldots b_{n-1}(t,z)$ and
$x_{ij}(t,z),\;0\leq i<\infty,\,i<j\leq i+n-1$ for some $n>1$,
so that $x_{i\,i+k}(t,z)$ is a polynomial $k$
for all sufficiently large $k$. Equation (~\ref{main_eq})
implies
\begin{eqnarray}
-zx_{0\,n-1}(t,z)'+a_{0\, n-1}(t,z)+x_{01}a_{1\,n-1}(t,z)
+\cdots +x_{0\,n-1}a_{n-1\, n-1}(t,z)
+{\bf x_{0\,n}}
\nonumber\\=
b_{1}(t,z)x_{0\, n-1}+b_{2}(t,z)x_{1\, n-1}+\cdots +b_{n-1}x_{n-2\, n-1}+b_{n}
,\nonumber\\
-zx_{i\,n+i-1}(t,z)'+a_{i\, n+i-1}(t,z)+x_{i\, i+1}a_{i+1\, n+i-1}(t,z)
\cdots x_{i\, n+i-1}a_{n+i-1\, n+i-1}(t,z)+{\bf x_{i\, n+i}}\nonumber\\=
x_{i-1\,n+i-1},\;i>0.\nonumber
\end{eqnarray}

As above we see that solving the $i$-th equation for $x_{i\, n+i}$ we obtain
\[x_{i\,n+i}(t,z)=q(i,t,z)+b_{n}(t,z)\]
for some polynomial $q(i,t,z)$ and all sufficiently large $i$. Again
the definition
of $\gtgl$ implies that
$x_{i\,n-i}(n+i,z)=0$ for all sufficiently large positive
integers $n$, and hence
$b_{n}(t,z)=-q(t-n,t,z)$.

 The described process shows that for any
$A\in p^{-1}_{g(t)}(f)$ there is at most one element of $N(z)$ converting
 it
to the canonical form.
It is easy to see that the
infinite matrix $(x_{ij})$ calculated above is an element
of $N(z)$.
 Proposition~\ref{action_group} follows  in the case of irrational
$g(t)$.

As to the rational $g(t)$ case, observe that, although the value of an
element
of the quotient ring at a point does not make sense, vanishing of
an element of the quotient ring at a point does make perfect sense in
our
case (see again property $(\ast)$ in the definition of $\gtgl$), and
so,
the same conversion  process completes the proof. $\Box$

\medskip

\begin{remark}
\label{help-pr-th}
 We point out another consequence of the conversion process.
For any $\lambda$, $\gtgl_{n}$ embeds naturally into $\bar{\gtgl}_{\lambda}$
by means of matrices with
 only finite number of non-zero columns and rows. This
gives rise to the embedding of the algebra of loops in the upper triangular
matrices and of the corresponding loop group. It is easy to see that this
 embedding is equivariant if $\lambda=N$ for all sufficiently large
positive integers $N$. In particular it induces an embedding of the quotients
$DO_{n}\hookrightarrow \Psi DS_{\lambda}$. Note that the last embedding is not
Poisson if $\lambda > n$.
\end{remark}

\subsubsection {Proof of Theorem 3.8}
{\bf A.} Filtrations of $\prod_{i\geq 1}\nc((z))$ and $p_{\lambda}^{-1}(f)$.
Recall that there is a filtration (c.f.(~\ref{filtrondiffop}))

\[W_{1}\subset W_{2}\subset\cdots\subset
Fun(\prod_{i=1}^{k} \nc((z))),\;\bigcup_{i\geq 1}W_{i}=
Fun(\prod_{i=1}^{k} \nc((z))).\]

Similarly one represents $p^{-1}(f)_{\lambda}$ as
 $p^{-1}(f)^{k}\times p^{-1}(f)^{-k},\;
k\geq 0$, where $p^{-1}(f)^{\pm k}$ is the set of matrices $f+(a_{ij})$, where
$a_{ij}=0$ if $j\leq i+k-1$ ($j> i+k-1$ respectively). Again let $j_{k}$ be the
projection on the first factor and set
$U_{k}=j_{k}^{\ast}(Fun(p^{-1}(f)^{k})),\; k\geq 1$.
 The result is the following
filtration
\[U_{1}\subset U_{2}\subset\cdots\subset
Fun(p^{-1}(f)),\;\bigcup_{i\geq 1}U_{i}=
Fun(p^{-1}(f)).\]
Consider the projection
 $\pi:\;p^{-1}(f)_{\lambda}\rightarrow \prod_{i\geq 1}\nc((z)).$
The group action is compatible with the filtration and therefore
\[\pi^{\ast}(W_{i})\subset U_{i}.\]

\medskip

{\bf B.} Proof of Theorem~\ref{main_th}. Let $\{,.,\}$ be the Poisson bracket
 on $(\hat{\gtgl}_{\lambda})^{\ast}$, $\{.,.\}^{\sim}_{\lambda}$ be the Poisson
bracket on $\prod_{i\geq 1}\nc((z))$ obtained as a result of hamiltonian
reduction,
$\{.,.\}_{\lambda}$ be the second Gelfand-Dickey structure on
$\prod_{i\geq 1}\nc((z))$. We have to show that
$\{.,.\}^{\sim}_{\lambda}=\{.,.\}_{\lambda}$ for all $\lambda$.

Let $f\in W_{i},\;g\in W_{j}$. Recall that
$\{f,g\}^{\sim}_{\lambda},\;\{f,g\}_{\lambda}$
are polynomials on $\lambda$.
 Therefore it is enough to prove that
\[\{f,g\}^{\sim}_{N}=\{f,g\}_{N}\]
for all sufficiently large $N$.

By definition,
\[\{U_{i},U_{j}\}\subset U_{i+j}.\]
Recall also ( see (~\ref{g_d_br-filtr}) ) that
\[\{W_{i},W_{j}\}_{\lambda}\subset W_{i+j}.\]
The last formula along with compatibility of $\pi$ with the filtrations
implies that
\[\{W_{i},W_{j}\}^{\sim}_{\lambda}\subset W_{i+j}.\]

 It follows that functions $\pi^{\ast} f,\;\pi^{\ast} g$ and their commutators
 are
uniquely determined by their restrictions to $\gtgl_{n}(z)\in
(\hat{\gtgl})^{\ast}$ (see Remark~\ref{help-pr-th}) for sufficiently
large $n$. Let $x\in \gtgl_{n}(z)\in
(\hat{\gtgl})^{\ast}$. One has

\begin{eqnarray}
\{f,g\}^{\sim}_{N}(\pi x)\nonumber\\
=\{\pi^{\ast} f,\pi^{\ast} g\}_{N}(x)\label{key_1}\\
=\{\pi^{\ast} f|_{\gtgl_{n}},\pi^{\ast} g|_{\gtgl_{n}}\}_{N}\nonumber\\
=\{f,g\}_{N}(\pi x)\label{key},
\end{eqnarray}
where (\ref{key_1}) follows from (~\ref{def_poiss_quot}),
  (~\ref{key}) follows Remark~\ref{help-pr-th}, Theorem~\ref{th-dr-sok}
and (~\ref{def_poiss_quot}).
Theorem~\ref{main_th} has been proved. $\Box$

\section{ Two more examples of deformation of Poisson
structures}

Algebra $\gtgl_{n}$ plays a universal role in mathematics for the reason
that almost  any  algebra maps into it. The algebra
 $\bar{\gtgl_{\lambda}}$ is expected to play a similar role in the deformation
theory. Here are a few examples.

\subsection { Deformation of Drinfeld-Sokolov reduction on orthogonal and
symplectic algebras}
In this section we construct 2 involutions: one on the space
of $\Psi DS_\lambda$, another on the algebra $\gtgl_{\lambda}$, such that
the Hamiltonian reduction sends a certain invariant subspace of one
to a certain invariant subspace of another.

1.{\em Gelfand--Dickey $\gtsp, \gtso$-brackets.}

To describe the Gelfand--Dickey structures corresponding to
the Lie algebras $\gtsp$ and $ \gtso$ we introduce the following involution $*$
on the set $\Psi DS_\lambda$ of pseudodifferential symbols:

\[ (\sum_{k=-\infty}^{0}u_k(z)D^{\lambda+k })^*=
\sum_{k=-\infty}^{0}(-1)^kD^{\lambda+k} u_k(z) \]

{\bf Definition.}
A pseudodifferential symbol $L$ is called {\em self-adjoint}
if $L^*=L$.

The set of self-adjoint pseudodifferential symbols $\Psi DS_\lambda^{SA}$
can be equipped  with the quadratic
Poisson structure in the same way as the set
$\Psi DS_\lambda$.
Having restricted the space of linear functionals to self-adjoint symbols one
can use the same Adler--Gelfand--Dickey formula  (~\ref{pole}).

We would like to emphasize  that the traditional definition
of the $\gtsp_{2n}$- ( $\gtso_{2n+1}$-) Gelfand--Dickey brackets
confines to the case of  self-adjoint (skew self-adjoint)
genuine differential operators of order $2n$ ($2n+1$, resp.).

2. {\em The simultaneous deformation of the
 algebras $\gtsp_{2n}, \gtso_{2n+1}$.}

Define the antiinvolution
$\sigma$ of $\gtsl_2$ to be the multiplication by $-1$.
Observe that $\sigma$ preserves the Casimir element $C=ef+fe+\frac{1}{2}h^{2}$.
Therefore $\sigma$ uniquely extends to an antiinvolution of $\gtgl_{\lambda}$,
which will also be denoted by the same letter $\sigma$. It is easy to see
that the eigenspace of $\sigma$ related to the eigenvalue $-1$ is a subalgebra.
Denote it by $\gtpo_{\lambda}$. The family of algebras
$\gtpo_{\lambda},
\;\lambda\in\nc$ is a deformation of both the families $\gtsp_{2n},\;
\gtso_{2n+1}$: if $\lambda=2n$ ($\lambda=2n+1$) the algebra
$\gtpo_{\lambda}$ contains $\gtsp_{2n}$ ($\gtso_{2n+1}$) as a quotient, see
{}~\cite{Fei88Alg}.

\begin{remark}
The algebra $\gtgl_{\lambda}$ is a direct sum of $\gtsl_2$-submodules
$\oplus_{i\geq 0} V_{2i+1}$. The involution $\sigma$ acts trivially
on the subspace $\oplus_{i\geq 0} V_{4i+1}$ and it acts by multiplication
by $-1$ on the subspace $\oplus_{i\geq 0} V_{4i+3}$.

A direct calculation shows that if the embedding of $\gtsl_{2}$
into $\gtgl_{n}$
is given by the image of $f$ (~\ref{ex_pr-sl2}):

\[f\mapsto \left( \begin{array}{cccccc}
0&0&0&\ldots&0&0\\
1&0&0&\ldots&0&0\\
0&1&0&\ldots&0&0\\
0&0&1&\ldots&0&0\\
.&.&.&.&.&.\\
0&0&0&\ldots&1&0 \end{array}
\right) \]
then the involution $\sigma$ acts on $\gtgl_{n}$ the following way. It
transposes a matrix with respect to the ``second diagonal" (not the main
diagonal, but the other one), and changes the sign of all entries that
are situated on every other shortened diagonal counting from the main one:
$\sigma(a_{i\ j})=(-1)^{i-j}a_{n-j-1\,n-i-1}$. Unlike the definition of
$\sigma$ we used previously, the latter
can not be carried over  to the case of  $\gtgl_{\lambda}$.
\end{remark}

One can extend the notions of trace, affinization, nilpotent subalgebra etc.
to the case of the algebra $\gtpo_{\lambda}$.

\begin{theorem}

The Hamiltonian reduction of the Kirillov--Kostant Poisson structure on
the algebra $\gtpo_{\lambda}$  results in the quadratic
Gelfand--Dickey structure on the space of pseudodifferential symbols
$\Psi DS_\lambda^{SA}$.
\end{theorem}

{\em Proof.}
This is an equivariant version of the main theorem ~\ref{main_th}.
Again, the result holds in virtue of  the finite dimensional analog
proved for $\gtsp_{2n}$ and $\gtso_{2n+1}$ by Drinfeld and Sokolov
(~\cite{DriSok84Alg}) and polynomial
 dependence on $\lambda$. $\Box$

\begin{remark}
Note that in the approach above it is possible to treat
the cases of self-adjoint and skew
self-adjoint operators on the same footing.
\end{remark}

\subsection {Deformation of the Toda lattice}

Recall the construction of the classical nonperiodic Toda lattice.
Let $E_{ij}$ be the matrix whose only non-zero entry is situated at the
intersection of the $i-$th row and $j-$th column and is equal to 1.
Let $\gtb_{n}\subset\gtgl_{n}$ be the subalgebra of upper-triangular
matrices and $\gtb_{n}^{-}\subset\gtgl_{n}$ the algebra of lower-triangular
matrices. Identify $\gtb_{n}^{\ast}$ with $\gtb_{n}^{-}$ by means of the trace.
 Set
\[\Lambda_{n}=\sum_{i=1}^{n-1}E_{i+1 \, i}\in\gtb_{n}^{\ast}.\]

 Let $\co_{\Lambda_n}$ be the orbit of
 $\Lambda_n$
in the coadjoint representation.

The dynamical system on $\co_{\Lambda_n}$ related to
the flow generated by the function
 $H_{2}(A)=Tr(A+\Theta)^{2}$ where
$\Theta=\sum_{i=1}^{n-1} i(n-i+1) E_{i \, i+1}$
 is called Toda lattice. Note that the element $\Theta$
can be replaced by any  linear combination of $E_{i\, i+1}$
with non-zero coefficients; our choice is
motivated by the generalization to the $\bar{\gtgl}_{\lambda}-$case, see below.
The function $H_{2}$ includes
in the family $H_{i},\;2\leq i\leq n$, where $H_{i}(A)=Tr(A+\Theta)^{i}$.
 Functions
$H_{i}$ are $\gtgl_{n}$-invariant, they Poisson commute as functions on
$\gtgl_{n}^{\ast}\approx\gtgl_{n}$ and, moreover, their restrictions
to $\gtb_{n}^{-}$ also Poisson commute, see ~\cite{Adl78Tra}.
 Calculations show that $dim \co_{\Lambda_n}=2n-2$. So, the Hamiltonian of the
Toda lattice has been included into the family of Poisson commuting functions,
the number of functions being equal to half the dimension of the phase  space.
This proves complete  integrability of the Toda lattice. (In fact one
also has to establish the independence of the functions, see ~\cite{Adl78Tra}.)

This all immediately carries over to the case of $\bar{\gtgl_{\lambda}},\;
\lambda\in\nc$:

Let $\gtb_{\lambda}\subset
\bar{\gtgl_{\lambda}}\subset\gtgl_{\infty}$ be the subalgebra of
 upper triangular matrices, $\gtb_{\lambda}^{-}\approx \gtb_{\lambda}^{\ast}$
subalgebra of lower triangular matrices.

\begin{proposition}
The subalgebra $\gtb_{\lambda}$ can be exponentiated to a Lie group.
\end{proposition}

{\em Proof.} Cf. Lemma~\ref{exp-nilp} and the Campbell-Hausdorf formula.

Set
\[\Lambda=\sum_{i=1}^{\infty}E_{i+1 i}\in \gtb_{\lambda}^{\ast}.\]
(We denoted this element $f$ in sect.3.)
Denote by $\co_{\Lambda}$ the corresponding orbit.

Take the element $\Theta=e \in \bar{\gtgl_\lambda}$.
Now we have infinitely many invariant
functions $H_{i}(A)=Tr(A+\Theta)^{i},\;i\geq 2$,
their restrictions to $\co_{\Lambda}$ are again independent and Poisson
commute (this is an obvious corollary of the corresponding finite dimensional
result). Therefore we have exhibited a family of infinite dimensional
integrable dynamical systems, ``containing'' classical Todda lattices
at the points $\lambda=n\in\nz$, and being approximated by the latter as
$n \rightarrow \infty$.


\begin{thebibliography}{7}

\bibitem{Adl78Tra}
M.~Adler, {\em On a trace functional for formal pseudo differential operators
  and the symplectic structure of the {K}orteweg-de{V}ries type equations},
  Inventiones Mathematicae {\bf 50} (1978/79), no.~3, 219--248.

\bibitem{Arn89Mat}
V.~I. Arnold, {\em Mathematical methods of classical mechanics},
Springer--Verlag, 2nd edition, 1989.

\bibitem{G-L-O}
M.~Golenishcheva-Kutuzova, D.~Lebedev and M.~Olshanetsky, {\em Between
$\hat{gl}(\infty)$ and $\hat{sl}_N$ affine algebras I. Geometrical actions.}
preprint ITEP-MO-94/1.

\bibitem{Japanese}
E.~Date, M.~Jimbo, M.~Kashiwara, and T.~Miwa, {\em Transformation group for
  soliton equations}, Publ. RIMS {\bf 18} (1982), 1077--1110.

\bibitem{Dick}
L.~A. Dickey, {\em Soliton equations and {H}amiltonian systems}, Advanced
  Series in Math. Physics, vol.~12, World Scientific, 1991.

\bibitem{DriSok84Alg}
V.~G. Drinfeld and V.~V. Sokolov, {\em Lie algebras and equations of
  {K}orteweg-de {V}ries type}, Current problems in mathematics (Moscow), Itogi
  Nauki i Tekhniki, vol.~24, Akad. Nauk SSSR, Vsesoyuz. Inst. Nauchn. i Tekhn.
  Inform., 1984, pp.~81--180 (Russian).

\bibitem{EnrKhorRad}
B.~Enriques, S.~Khoroshkin, A.~Radul, A.~Rosly, and V.~Rubtsov, {\em
  Poisson-{L}ie aspects of classical {W}-algebras}, preprint, Ecole
  Polytechnique, 1993.

\bibitem{Fei88Alg}
B.~L. Feigin, {\em Lie algebras ${gl}(\lambda)$ and cohomology of
  a {L}ie algebra of differential operators}, Russian Mathematical Surveys {\bf
  43} (1988), no.~2, 169--170.

\bibitem{Fre}
I.~B.~Frenkel, {\em Orbital theory for affine Lie algebras}, Invent. Math.
{\bf 77} (1984), 301-352.

\bibitem{GelDick78}
I.~M. Gelfand and L.~A. Dickey, {\em A family of {H}amiltonian structures
  associated with nonlinear integrable differential equations}, preprint, IPM
  AN SSSR, Moscow, 1978.


\bibitem{KP}
V.~G. Kac and D.~H. Peterson, {\em Spin and wedge representations of
  infinite-dimensional {L}ie algebras and groups}, Proc. Nat. Ad. Sci. USA {\bf
  78} (1981), 3308--3312.

\bibitem{KhesZakh93Poi}
B.~A. Khesin and I.~S. Zakharevich, {\em Poisson-Lie group
of pseudodifferential symbols and fractional
  {KP}-{KdV} hierarchies}, C.R.Acad.Sci. {\bf 316} (1993), 621--626;
{\em Poisson-Lie group of pseudodifferential symbols}, preprint
hep-th 9312088, to appear in Comm. Math. Phys.



\bibitem{Kos}
B.~Kostant, {\em The principal three-dimensional
subgroup and the Betti numbers of a complex simple Lie group
}, Amer. J. Math. {\bf 81} (1959), 973-1032.

\bibitem{Kos2}
B.~Kostant, {\em Lie group representations
on polynomial rings}, Bull. Amer. Math. Soc. {\bf 61} (1967) n.4.

\bibitem{KhesKra91Cen}
O.~S. Kravchenko and B.~A. Khesin, {\em Central extension of the algebra of
  pseudodifferential symbols}, Funct. Anal. Appl. {\bf 25} (1991), no.~2,
  83--85.

\bibitem{KupWil81Mod}
B.~A. Kupershmidt and G.~Wilson, {\em Modifying {L}ax equations and the
  second {H}amiltonian structure}, Inventiones Mathematicae {\bf 62} (1981),
  no.~3, 403--436.


\bibitem{Mal_F_F}
F.~Malikov, B.~Feigin, D.~Fuchs {\em Singular vectors in Verma modules
over Kac-Moody algebras}, Funkc.Anal. i ego Pril. {\bf 20} (1986) 2 25-37

\bibitem{Mar-Wei}
J.~E.~Marsden and A.~Weinstein, {\em Reduction of symplectic manifolds with
symmetries}, Rep. Math. Phys. {\bf 5} (1974), 121-130.

\bibitem{Rei-Sem}
A.~Reiman and M.~Semenov-Tian-Shansky, {\em Lie algebras and nonlinear partial
differential equations}, Sov. Math. Doklady
{\bf 21} (1980), 630-636.


\bibitem{We}
 A.~Weinstein, {\em Local structure of Poisson manifolds},
 J.Diff. Geom. {\bf 18} (1983), no.~3, 523-558.



\end{thebibliography}
\end{document}